
\magnification=1200
\baselineskip=20pt
\def\lsim{<\kern-2.5ex\lower0.85ex\hbox{$\sim$}\ }
\def\rsim{>\kern-2.5ex\lower0.85ex\hbox{$\sim$}\ }
\def\sqr#1#2{{\vcenter{\vbox{\hrule height.#2pt
         \hbox{\vrule width.#2pt height#1pt \kern#1pt
            \vrule width.#2pt}
         \hbox height.#2pt}}}}

\def\bull{\vrule  height .9ex width .8ex depth -.1ex}
\overfullrule=0pt
\centerline{\bf TRIPLE PRODUCTS AND YANG--BAXTER EQUATION (II):}
\centerline{\bf ORTHOGONAL AND SYMPLECTIC TERNARY SYSTEMS}
\vskip 1cm
\centerline{by}
\vskip 1cm
\centerline{Susumu Okubo}
\centerline{Department of Physics and Astronomy}
\centerline{University of Rochester}
\centerline{Rochester, NY 14627, USA}
\vskip 3 cm
\noindent {\bf \underbar{Abstract}}

We generalize the result of the preceeding paper and solve the Yang--Baxter
equation in terms of triple systems called orthogonal and symplectic
ternary systems.  In this way, we found several other new solutions.

\vskip 2cm
\noindent {\underbar{PAC}: 03.65Fd. 02.90.+p
\vfil\eject
\noindent{\bf 1. \underbar{Introduction and Summary of Results}}

Let $V$ be a $N$-dimensional vector space with a bi-linear non-degenerate
form (or inner product) $<x \vert y>$ for $x, y \ \epsilon\ V$.  Let
 $e_1, e_2, \dots, e_N$ be a basis of $V$ and set
$$<e_j \vert e_k> \ = g_{jk} \eqno(1.1)$$
with its inverse $g^{jk}$ satisfying
$$g^{jk} g_{k \ell} = \delta^j_\ell \quad . \eqno(1.2)$$
We raise or lower indices, as usual, by $g^{jk}$ or $g_{jk}$.  For example,
we set
$$e^j = g^{jk} e_k \eqno(1.3)$$
so that we have
$$<e^j \vert e_k >\ = \delta^j_k \eqno(1.4)$$
as well as
$$e_j <e^j \vert x>\ =\ < x \vert e_j>e^j = x \quad . \eqno(1.5)$$
In the preceeding paper$^{1)}$ which we refer to hereafter as I, we have
rewritten the Yang--Baxter (Y--B) equation
$$\eqalign{R^{b^\prime a^\prime}_{a_1 b_1} (\theta)
&R^{c^\prime a_2}_{a^\prime c_1} (\theta^\prime)
R^{c_2 b_2}_{b^\prime c^\prime} (\theta^{\prime \prime})\cr
&= R^{c^\prime b^\prime}_{b_1 c_1} (\theta^{\prime \prime})
R^{c_2 a^\prime}_{a_1 c^\prime} (\theta^\prime)
R^{b_2 a_2}_{a^\prime b^\prime} (\theta)\cr}\eqno(1.6)$$
with
$$\theta^\prime = \theta + \theta^{\prime \prime} \eqno(1.7)$$
as a triple product equation
$$\eqalign{[v,[u,e_j,&z]_{\theta^\prime},
[e^j,x,y]_{\theta}]_{\theta^{\prime \prime}}\cr
&= [u,[v,e_j,x]_{\theta^\prime},[e^j,z,y]_{\theta^{\prime \prime}}
]_\theta \quad ,\cr}\eqno(1.8)$$
provided that the scattering matrix element $R^{ab}_{cd} (\theta)$ satisfies
the symmetry condition
$$R^{ab}_{cd} (\theta) = R^{ba}_{dc} (\theta) \eqno(1.9)$$
or equivalently
$$< u \vert [z,x,y]_\theta>\ = \ <z \vert [u,y,x]_\theta >\quad .
\eqno(1.10)$$
Here, the $\theta$-dependent triple product $[x,y,z]_\theta$ has been
defined by
$$R^{ab}_{cd} (\theta) = \ <e^a \vert [e^b , e_c , e_d]_\theta> \quad .
\eqno(1.11)$$
The more general case without assuming Eq. (1.9) will be discussed in
section 6.  In I, we have solved the Y--B equation for two cases of
$N= 4$ and 8, corresponding to quaternionic and octonionic triple products.
 The purpose of this note is to generalize the method for more general
cases of any orthogonal and symplectic ternary systems satisfying a
condition to be specified shortly.  To be definite, we shall first give
axioms for these systems below.  Suppose that the vector space $V$
possesses a \underbar{$\theta$-independent} triple product
$$xyz\ :\ V \otimes V \otimes V \rightarrow V \eqno(1.12)$$
as well as the non-degenerate bi-linear form $<x \vert y>$.  Let
$\varepsilon$ be a constant assuming value of either $\varepsilon =
+1$ or $\varepsilon = -1$.  Our fundamental ansatz is then that they
satisfy axioms:
\itemitem{(i)} $<y \vert x>\ = \varepsilon < x \vert y>$ \hfill
(1.13a)
\itemitem{(ii)} $xyz + \varepsilon\   yxz = 0$ \hfill
(1.13b)
\itemitem{(iii)} $xyz + \varepsilon\   xzy = 2 \lambda <y \vert z>x -
\lambda <x \vert y>z - \lambda <z \vert x>y$ \hfill (1.13c)
\itemitem{(iv)} $uv(xyz) = (uvx)yz + x(uvy)z +
 xy(uvz)$ \hfill (1.13d)
\itemitem{(v)} $<uvx \vert y>\ = - <x \vert uvy>$ \hfill (1.13e)

\noindent for $u,v,x,y,z\ \epsilon \ V$, where $\lambda$ in Eq. (1.13c) is
a constant.  Then, the case of $\epsilon = +1$ defines the orthogonal
ternary system (OTS) as in I, while the other case of $\epsilon = -1$ is
called by Yamaguchi and Asano$^{2)}$ to be the symplectic ternary system
(STS).  Both OTS and STS may be regarded as special cases of more general
triple systems discussed by many authors$^{3)-9)}$, whose studies will be
left, however, in the future.

Before going into further details, we note that the last postulate Eq.
(1.13e) is actually a consequence of other postulates Eqs.
 (1.13a)--(1.13d), provided that we have $\lambda \not= 0$ and Dim $V \geq
2$ for $\varepsilon = 1$.  However, since we consider sometime the special
case of $\lambda =0$, we added it as an extra postulate here.  To show it,
we first introduce the notion of a derivation
$$D\  :\ V \rightarrow V \eqno(1.14)$$
of the triple system to be a linear transformation in $V$ satisfying
$$D(xyz) = (Dx)yz + x(Dy)z + xy(Dz) \quad . \eqno(1.15)$$
Applying $D$ to both sides of Eq. (1.13c), we find then an identity
$$\eqalign{2 \lambda \{ <Dy \vert z>\ &+ <y \vert Dz> \}x\cr
&- \lambda \{<z \vert Dx>\ + \ <Dz \vert x>\}y\ -
\lambda \{ <Dx \vert y> \ + \ <x \vert Dy>\}z
 = 0 \quad .\cr}$$
Suppose $\lambda \not= 0$, and set $z = x$.  For $\varepsilon = -1$, this
immediatley gives
$$<x \vert Dy>\ = - <Dx \vert y> \eqno(1.16)$$
as has already been observed by Yamaguchi and Asano$^{2)}$.  For the other
case of $\varepsilon = +1$, Eq. (1.16) will also follow, provided that we
have Dim $V \geq 2$ which we will assume hereafter.  Next, if we introduce
 the left multiplication operator $L_{x,y}\ :\ V \rightarrow V$ by
$$L_{x,y} z = xyz$$
then Eq. (1.13d) implies that $D = L_{u,v}$ is a derivation of the system
so that Eq. (1.13e) will follow readily from Eq. (1.16).  In passing, Eq.
(1.13d) can be rewritten as a Lie equation
$$[L_{u,v}, L_{x,y}] = L_{uvx,y} +
 L_{x,uvy} = -L_{xyu,v} - L_{u,xyv} \eqno(1.17)$$
which is equivalent to
$$\eqalign{uv(xyz) - xy(uvz) &= (uvx)yz + x(uvy)z\cr
&= - (xyu)vz - u(xyv)z \quad .\cr}\eqno(1.18)$$
For the octonionic triple product corresponding to $\varepsilon = 1$
and $N= \ {\rm Dim}\ V = 8$, Eq. (1.17) defines a so(8) Lie
algebra, though $V$ is the 8--dimensional module of the so(7).

Next, we introduce the second triple product
$$[x,y,z]\ :\ V \otimes V \otimes V \rightarrow V \eqno(1.19)$$
by
$$[x,y,z] = xyz - \lambda <y \vert z>x + \lambda <z \vert x>y \quad .
\eqno(1.20)$$
Then, Eqs. (1.13b), (1.13c) and (1.13e) can be restated as the statement
that both $[x,y,z]$ and $<w \vert [x,y,z]>$ are  totally
antisymmetric for $\varepsilon = 1$ and totally symmetric for $\varepsilon =
-1$, respectively, with respect to
 3 variables $x,\ y,\ {\rm and}\ z$ or
 4-variables $x,y,z,\ {\rm and}\ w$.
However, the derivation property Eq. (1.13d) becomes rather complicated in
terms of $[x,y,z]$.  We will profitably utilize, in this note, both notations,
alternatively depending upon situations.

Let $e_j (j = 1,2,\dots,N)$ be a basis of $V$.  Then, $g_{jk}$ defined by
Eq. (1.1) satisfies now
$$g_{kj} = \varepsilon \ g_{jk} \quad , \eqno(1.20)$$
so that we have
$$e_j \otimes e^j = \varepsilon \ e^j \otimes e_j \quad , \eqno(1.21)$$
$$[x,e_j , e^j] = 0 \quad . \eqno(1.22)$$
Moreover, in view of Eqs. (1.5) and (1.13e), we can readily see the
validity of
$$xye^j \otimes e_j = - e^j \otimes xye_j = - \varepsilon\
e_j \otimes xye^j \eqno(1.23)$$
which will be used often in what follows.

We organize our paper as follows.   In section 2, we will study further
consequences of both OTS and STS.  Especially, we will first show that
another triple product $x \cdot y \cdot z$ given by
$$x \cdot y \cdot z = (xye^j) z e_j - {1 \over 3}\
\lambda (\epsilon N-7)xyz \eqno(1.24)$$
defines Lie$^{10)}$ and anti-Lie$^{11)}$ triple products for
$\varepsilon = 1 \ {\rm and}\ \varepsilon = -1$, respectively.  In section
3, we will solve the Y--B equation in a form of
$$\eqalign{[z,x,y]_\theta = &P(\theta) xyz + Q(\theta) <x \vert
y>z\cr
&+ R(\theta) <z \vert x >y + S(\theta)<y \vert z>x \cr}\eqno(1.25)$$
for some functions $P(\theta), \ Q(\theta),\ R(\theta),\ {\rm and}\
S(\theta)$ to be determined, assuming that OTS or STS satisfies the
additional condition of
$$x \cdot y \cdot z = 0 \eqno(1.26)$$
identically.  Note the change of orders of variables $x,y,\ {\rm and}\
z$ in $[z,x,y]_\theta$ and $P(\theta) xyz$ in Eq. (1.25).  This is
necessary in order to accomodate the symmetry condition Eq. (1.10)
 so that we need only solve Eq. (1.8).  Rewriting
$[z,x,y]_\theta$ as
$$\eqalign{[z,x,y]_\theta = &P(\theta) [x,y,z] + A(\theta) <x \vert y>z\cr
&+ B(\theta) <z \vert x>y + C(\theta) <y \vert z>x\cr}\eqno(1.27)$$
with
$$\eqalignno{A(\theta) &= Q(\theta)\quad ,&(1.28a)\cr
B(\theta) &= R(\theta) - \lambda P(\theta) \quad , &(1.28b)\cr
C(\theta) &= S(\theta) + \lambda P(\theta)\quad ,&(1.28c)\cr}$$
the solution of the Y--B equation is found in section 3 to be
$$\eqalignno{A(\theta) &= \big\{ \lambda - {2 a \lambda \over
2(a-\lambda) + b \theta}\big\} \ P(\theta) \quad ,&(1.29a)\cr
B(\theta) &= (a - \lambda + b \theta) P(\theta)\quad , &(1.29b)\cr
C(\theta) &= \big( - \lambda - {2 a \lambda \over b \theta}\big)
\ P(\theta) \quad ,&(1.29c)\cr}$$
where we have set for simplicity
$$a = {1 \over 6}\ \lambda (4 - \epsilon N) \eqno(1.30)$$
while $P (\theta)$ is an undetermined function of $\theta$, and $b$ is an
arbitrary constant.

In section 4, we will discuss various OTS and STS satisfying the condition
Eq. (1.26), i.e. $x \cdot y \cdot z = 0$.  It will be shown there that for
$\varepsilon = 1$ (OTS), both octonionic and Malcev triple products with
$N=8$ and $N=7$, respectively satisfy the condition.  Especially, for the
former, the solution Eqs. (1.29) will reproduce the result of I with
$\lambda = -3 \beta = 3$.  However, the quaternionic triple product with
$\varepsilon N = 4$ whose solution has been given in I does not satisfy
$x \cdot y \cdot z = 0$.  With respect to the STS case of $\varepsilon =
-1$, we have found six solutions with $N=2,\ 4,\ 14, \ 20,\ 32,\ {\rm and}
\ 56$.  They are intimately related to the Lie algebras $A_2,\  G_2,\
F_4,\ E_6,\ E_7,\ {\rm and}\ E_8$.  Especially, the last solution of $N=56$
corresponds to the celebrated Freudenthal's triple system$^{12)}$.  Also,
for the simplest case of $N=2$, we can find more general solutions which
are either constant or of trigonometric type, as will be studied in section
5.  Finally, we will show in section 6 how to rewrite Eq. (1.6) as a triple
product equation without assuming Eqs. (1.9) or (1.10).
\medskip
\noindent {\bf 2. \underbar{Orthogonal and Symplectic Ternary Systems}}

In this section, we will study various consequences of OTS and STS, which
will be needed for the solution of the Y--B equation to be given in section
3.

First, we note that $xyz$ given by
$$xyz = \lambda <y \vert z>x - \lambda <z \vert x>y$$
or equivalently
$$[x,y,z] = 0$$
satisfies all axioms Eqs. (1.13b)--(1.13e) of OTS and STS when we note Eq.
(1.13a).  We call such a case to be trivial.

We now prove the following lemma.
\eject
\smallskip
\noindent \underbar{Lemma 1}

Let $V$ be either OTS or STS.  Then, we have
\itemitem{(i)} $e_je^jx = 0 \quad , \quad <e_j \vert xye^j>\ = 0$
\hfill (2.1a)
\itemitem{(ii)} $xe_je^j = \lambda (\epsilon N-1)x$\hfill (2.1b)
\itemitem{(iii)} $<u \vert xvy>\ = \ <v \vert yux>$\hfill (2.1c)
\itemitem{(iv)} $<z \vert xye^j>e_j = - \varepsilon \ xyz$\hfill (2.1d)
\itemitem{    } $<x \vert e_j yz>e^j = - zxy$ \hfill (2.1e)
\itemitem{(v)} $<xye^j \vert zue_j>\ = -<(xye^j)e_j z \vert u>$
\itemitem{   } $\qquad =\ <u \vert (xye^j) ze_j>\ - 3 \lambda
<u \vert xyz>$\hfill (2.1f)
\itemitem{(vi)} $(xye^j)ze_j = ze_j(xye^j) = (xe^jy)e_jz =
- \varepsilon (xye^j)e_jz + 3 \lambda xyz$\hfill (2.1g)
\itemitem{(vi)} $<v \vert (xye^j)zu>e_j = \epsilon \ xy(uvz)\quad .$
\hfill (2.1h)
\smallskip
\noindent \underbar{Proof}

 Noting Eqs. (1.13b) and (1.21), we calculate
$$e_je^jx = - \varepsilon \ e^je_jx = - \varepsilon \varepsilon \ e_j
e^j x = -e_j e^j x=0$$
which proves the 1st relation in Eq. (2.1a).  The 2nd relation can be
similarly obtained when we use Eqs. (1.13a) and (1.13e).
   Next, we compute
$$\eqalign{x e_j e^j + \varepsilon \ x e^j e_j &= 2 \lambda <e_j \vert e^j>x -
 \lambda <x \vert e_j>e^j - \lambda <e^j \vert x>e_j\cr
&= 2 \lambda\  \varepsilon\ N \ x - \lambda x -
\lambda x = 2 \lambda (\epsilon N-1)x\cr}$$
\noindent from Eqs. (1.13a),
 (1.13c), (1.4) and (1.5).  On the other side, we have
$$\varepsilon \ x\ e^je_j = \varepsilon \varepsilon \ x\ e_je^j =
x e_j e^j$$
in view of Eq. (1.21).  This proves Eq. (2.1b).  As for (iii), we rewrite
$$\eqalign{<u \vert xvy>\ &=\  <u \vert [x,v,y]>\ +\  \lambda <v \vert y>
<u \vert x> \ -\  \lambda <y \vert x><u \vert v> \quad ,\cr
<v \vert yux>\ &=\  <v \vert [y,u,x]>\ +\  \lambda <u \vert x>
<v \vert y> \ -\  \lambda <x \vert y><v \vert u> \cr}$$
and note that $<u \vert [x,v,y]>$ is totally symmetric for
$\varepsilon = 1$ and totally antisymmetric for $\varepsilon = -1$,
respectively.  Comparing both, this gives Eq. (2.1c).

Next, Eq. (2.1d) is a immediate consequence of Eqs. (1.13e), (1.13a) and
(1.5), while Eq. (2.1e) follows then from Eqs. (2.1d)
 and (2.1c) when we rewrite it
$$<x \vert e_jyz>e^j = \ <y \vert zxe_j>e^j = \varepsilon
<y \vert zxe^j>e_j = -zxy \quad .$$

Replacing $v \rightarrow u \rightarrow xye^j,\  x \rightarrow z \ {\rm and}
\ y \rightarrow e_j$ in Eq. (2.1c), we find
$$<xye^j \vert zue_j>\ = \ <u \vert e_j (xye^j)z>\ = - \varepsilon
<u \vert (xye^j)e_jz>\  = - <(xye^j) e_j z \vert u>$$
which gives the 1st relation in Eq. (2.1d).  Then, the last relation in Eq.
(2.1d) follows from Eq. (2.1g) to be proved below.  Now Eq. (1.23) readily
gives
$$(xye^j)ze_j = - \varepsilon\ e_jz(xye^j) = ze_j(xye^j)$$
which proves the 1st relation in Eq. (2.1g).  In order to show the rest of
equations, we calculate
$$\eqalign{(xe^jy)e_jz &= \{-\varepsilon\ xye^j + 2\lambda<e^j \vert y>x -
 \lambda <x \vert e^j>y - \lambda <y \vert x>e^j \} e_j z\cr
&= - \varepsilon (xye^j)e_jz + 2 \lambda <e^j \vert y>xe_jz -
\lambda <x \vert e^j>ye_jz - \lambda <y \vert x>e^j e_j z\cr
&= - \varepsilon (xye^j)e_jz + 2 \lambda
\  xyz - \lambda \epsilon\ yxz\cr
&= - \varepsilon (xye^j)e_jz + 3 \lambda
\  xyz\cr
&= - \varepsilon \{- \varepsilon (xye^j)ze_j
 + 2 \lambda
<e_j \vert z>xye^j\cr
&\quad - \lambda <xye^j \vert e_j>z - \lambda <z \vert xy e^j>e_j\} +
3 \lambda \ xyz\cr
&= (xye^j)ze_j - 2 \lambda
\  xyz - \lambda \ xyz + 3 \lambda \ xyz\cr
&= (xye^j)ze_j\cr}$$
when we note Eqs. (1.13b), (1.13c), (1.5), (2.1a) and (2.1d).
Finally, from Eqs. (2.1c) and (1.13e), we find
$$\eqalign{<v \vert (xye^j)zu>e_j &= \ <z \vert uv(xye^j)>e_j = -
<uvz \vert xye^j>e_j\cr
&= \ <xy (uvz)\vert e^j>  e_j = \varepsilon\ xy(uvz)\cr}$$
which is Eq. (2.1h).  This completes the proof of the Lemma 1. $\ \bull$
\vfil\eject
\smallskip
\noindent \underbar{Proposition 1}

The new triple product defined by
$$\eqalign{x \cdot y \cdot z &= (xye^j)ze_j - {1 \over 3}\
\lambda (\epsilon N-7)xyz \quad ,\cr
&= - (xye_j)e^jz -
{1 \over 3}\ \lambda (\epsilon N-16) xyz \cr}\eqno(2.2)$$
satisfies the following relations:
\smallskip
\itemitem{(i)} $x \cdot y \cdot z = - \varepsilon\ y \cdot x \cdot z
\quad ,$
\hfill (2.3a)
\itemitem{(ii)} $x \cdot y \cdot z + y \cdot z \cdot x + z \cdot x \cdot
y = 0 \quad ,$ \hfill (2.3b)
\itemitem{(iii)} $u \cdot v \cdot(x \cdot y \cdot z)
 = ( u \cdot v \cdot x)\cdot y \cdot z + x\cdot (u \cdot v \cdot
y)\cdot z + x \cdot y \cdot (u \cdot
v \cdot z)\quad ,$  \hfill (2.3c)
\itemitem{(iv)} $uv (x \cdot y \cdot z) =
(uvx) \cdot y \cdot z + x \cdot (uvy) \cdot z +
x \cdot y \cdot (uvz)\quad ,$ \hfill (2.3d)
\itemitem{(v)} $u \cdot v \cdot (xyz) =
(u \cdot v \cdot x)yz  + x(u \cdot v \cdot
y)z + xy(u \cdot v \cdot z)\quad ,$  \hfill (2.3e)
\itemitem{(vi)} $<u \cdot v \cdot x \vert y>\
= -< x \vert u \cdot v \cdot y>\
= - <v \vert  y \cdot x \cdot
u> \quad .$  \hfill (2.3f)
\smallskip
\noindent Especially Eqs. (2.3a)--(2.3c)
 imply that it defines a Lie$^{10)}$ or
anti-Lie$^{11)}$ triple system, respectively, for $\varepsilon = 1$ or
$\varepsilon = -1$.
\smallskip
\noindent \underbar{Proof}

The first relation Eq. (2.3a) is a immediate consequence of Eq. (1.13b).

Next by the derivation relation Eq. (1.13d), we calculate
$$\eqalign{ze_j(xye^j) &= (ze_jx)ye^j + x(ze_jy)e^j +
xy(ze_je^j)\cr
&= (ze_jx)ye^j - \varepsilon(ze_jy)xe^j + \lambda (\epsilon
N-1) xyz \quad .\cr}\eqno(2.4)$$
Moreover, we continue
$$\eqalign{(ze_jx)ye^j &=
 \{- \epsilon\  zxe_j + 2 \lambda <e_j \vert x>z - \lambda <z
\vert e_j>x - \lambda <x \vert z>e_j \}ye^j\cr
&= - \varepsilon (zxe_j)ye^j + 2 \lambda \ \varepsilon \ zyx -
 \lambda xyz +  \lambda \varepsilon <x \vert z>ye_je^j\cr
&= -(zxe^j)ye_j - 2 \lambda \ yzx - \lambda xyz +
\lambda^2 <z \vert x>(\epsilon N-1)y\cr
&= -(zxe^j)ye_j - 3 \lambda [x,y,z] -
\lambda^2 <y \vert z>x \cr
&\quad +2 \lambda^2 <x \vert y>z + \lambda^2 (\epsilon N-2)<z \vert x>y\cr}
$$
after some calculation.  Then, Eq. (2.4) together with Eq. (2.1g) leads to
$$\eqalign{(xye^j)ze_j &+ (zxe^j)ye_j + (yze^j)xe_j\cr
&= \lambda (\epsilon N-7)[x,y,z] = {1 \over 3}\
\lambda (\epsilon N-7)(xyz + yzx + zxy)\cr}$$
which gives Eq. (2.3b).

In order to prove Eqs. (2.3d)--(2.3f), we first set
$$x*y*z = (xye^j) ze_j = - (xye_j)e^jz + 3 \lambda xyz \quad , \eqno(2.5)$$
and calculate
$$\eqalign{uv(x*y*z) &- (uvx)*y*z - x*(uvy)*z - x*y*(uvz)\cr
&= \{xy(uve^j)\}ze_j + (xye^j)z(uve_j)\cr}$$
\noindent from Eq. (1.13d).
 However, the right side of this relation is identically
zero because of Eq. (1.23)
i.e. $(uve^j) \otimes e_j + e^j \otimes (uve_j) = 0$,
 so that we find
$$uv(x*y*z) = (uvx)*y*z + x*(uvy)*z + x*y*(uvz)\quad. \eqno(2.6)$$
Then, Eq. (2.3d) follows readily from Eq. (2.6) and
$$x \cdot y \cdot z = x*y*z - {1 \over 3}\ \lambda
(\epsilon N-7)xyz \quad .\eqno(2.7)$$
Next, we note
$$(uve^j)e_j(xyz) = \{(uve^j)e_jx\} yz + x\{ (uve^j)e_jy\}z +
 xy\{(uve^j)e_jz\} \quad .$$
Together with Eq. (2.5), this gives
$$u*v*(xyz) = (u*v*x)yz + x (u*v*y)z + xy(u*v*z) \eqno(2.8)$$
and hence Eq. (2.3e).  Similarly, Eq. (2.3c) is a consequence of
$$u*v*(x*y*z) = (u*v*x)*y*z + x*(u*v*y)*z + x*y*(u*v*z) \eqno(2.9)$$
which results from Eq. (1.13d) as well as
$$(x*y*e^j) \otimes e_j = - e^j \otimes (x*y*e_j) \quad , \eqno(2.10a)$$
or equivalently
$$<x*y*u \vert v>\  = - <u \vert x*y*v> \eqno(2.10b)$$
which is the analogue of Eq. (1.23).  Finally, we can verify Eq. (2.3f)
directly.  We may note that if $\lambda \not= 0$, then $<u \cdot v\cdot
x \vert y>\  = - <x \vert u \cdot v \cdot y>$ is a simple consequence of Eq.
(1.16), since $D = L^*_{x,y}$ defined by
$$L^*_{x,y} z = x \cdot y \cdot z$$
is a derivation of the original OTS or STS because of Eq. (2.3e).  At any
rate, these complete the proof of the Proposition 1.
\smallskip
\noindent \underbar{Proposition 2}

We have
$$\eqalign{2 (xye^j)(zue_j)v &= \varepsilon (x \cdot y \cdot z)
uv - (x \cdot y \cdot u)zv\cr
&\quad + {1 \over 3}\ \varepsilon \lambda (\epsilon N-16) (xyz)uv -
 {1 \over 3}\ \lambda (\epsilon N-16) (xyu)zv \quad .\cr}\eqno(2.11)$$
\smallskip
\noindent \underbar{Corollary 1}

We have
\itemitem{(i)} $\varepsilon (xyz)uv - (xyu)zv + \varepsilon
(zux)yv - (zuy)xv = 0$
\itemitem{(ii)} $\varepsilon (x \cdot y \cdot z) uv - (x \cdot y \cdot u)zv
+ \varepsilon (z \cdot u \cdot x)yv - (z \cdot u \cdot y)xv = 0$

\smallskip
\noindent \underbar{Proof}

We first rewrite Eq. (1.13d) as
$$\eqalign{(uvx)yz &+ (uvy)zx + (uvz)xy - uv[x,y,z]\cr
&= \lambda <y \vert z>uvx + \lambda <z \vert x>uvy + \lambda <x \vert y>
uvz\cr
&\quad - \lambda <y \vert uvz>x - \lambda <z \vert uvx>y - \lambda <x \vert
uvy>z \quad , \cr}\eqno(2.12)$$
and let $x \rightarrow u \rightarrow xye^j,\ y \rightarrow e_j, \
{\rm and}\ v \leftrightarrow z$ in Eq. (2.12) to find
\smallskip
$$\eqalign{[(xye^j)&zu]e_jv + [(xye^j)ze_j]vu + [(xye^j)zv]u e_j -
(xye^j)z[u,e_j,v]\cr
&= \lambda <e_j \vert v>(xye^j) zu + \lambda <v \vert u>(xye^j)ze_j +
 \lambda <u \vert e_j>(xye^j)zv\cr
&\quad - \lambda <e_j \vert (xye^j) zv>u\  -\
 \lambda <v \vert (xye^j)zu>e_j -
 \lambda <u \vert (xye^j)ze_j>v \quad .\cr}\eqno(2.13)$$

We now rewrite 4-terms in the left side of Eq. (2.13) respectively as
$$\eqalign{[(xye^j&)zu]e_jv\cr
&= (xye^j)(zue_j)v - 2 \lambda (xyu)zv\cr
&\quad  + \lambda \varepsilon (xyz)uv +
 \lambda <z \vert u>(xye^j)e_jv \quad ,\cr}\eqno(2.14a)$$
\smallskip
$$\eqalign{[(xye^j&)ze_j]vu\cr
&= vu[(xye^j)ze_j] - 2 \lambda <u \vert (xye^j)ze_j>v\cr
&\quad  + \lambda \varepsilon <v \vert (xye^j)ze_j>u
 + \lambda \varepsilon  <u \vert v>(xye^j)ze_j \quad ,\cr}\eqno(2.14b)$$
\smallskip
$$\eqalign{[(xye^j&)zv]ue_j\cr
&= - \varepsilon (xye^j)(zve_j)u + 2 \lambda
\varepsilon  (xyv)zu
 - \lambda (xyz)vu\cr
&\quad + 2 \lambda (xyu)zv - \lambda xy (vuz)\cr
&\quad  - \lambda <v \vert z>(xye^j)e_ju
+ \lambda \varepsilon <v \vert (xye^j)ze_j>u
 \quad ,\cr}\eqno(2.14c)$$
\smallskip
$$\eqalign{-(xy&e^j)z[u,e_j,v]\cr
&= -(xye^j)(uve_j)z - 2 \lambda \varepsilon\
 xy(uvz) + \lambda \varepsilon\ uv (xyz)\cr
&\quad  - \lambda \varepsilon (xyv)zu +
 \lambda (xyu)zv - \lambda \varepsilon
<uve_j \vert xye^j>z \quad ,\cr}\eqno(2.14d)$$
in the following ways.

First consider Eq. (2.14a).  In view of Eq. (1.23), we calculate
\smallskip
$$\eqalign{[(xye^j)zu]e_jv &= -(e^jzu)(xye_j)v = \varepsilon (xye_j)
(e^jzu)v\cr
&= (xye^j)(e_jzu)v = - \varepsilon (xye^j)(ze_ju)v\cr
&= -\varepsilon (xye^j)\{ - \varepsilon \ zue_j + 2 \lambda <e_j \vert u>z -
\lambda <z \vert e_j>u - \lambda <u \vert z >e_j\}v\quad ,\cr}$$
\smallskip
\noindent which together
 with Eq. (1.5) leads to Eq. (2.14a).  The next relation Eq.
(2.14b) is a direct consequence of Eqs. (1.13b) and (1.13c).  As for Eq.
(2.14c), we first note
$$[(xye^j)zv]ue_j = - (e^jzv)u(xye_j)$$
by Eq. (1.23).  Then, Eq. (2.14c) follows after some calculations in  a
similar way.  The same remark also applies for the derivation of Eq.
(2.14d).

The right side of Eq. (2.13) can be readily computed from Eqs. (1.5) and
(2.1h) to be
$$\eqalign{\lambda \epsilon (xyv)zu &+
 \lambda (xyu)zv - \lambda \varepsilon\ xy(uvz)
+ \lambda \varepsilon <u \vert v>(xye^j)ze_j\cr
& + \lambda \varepsilon
<v \vert (xye^j)ze_j>u - \lambda <u \vert (xye^j)ze_j>v \quad .
\cr}\eqno(2.15)$$
\noindent From Eqs. (2.14) and (2.15),
 we can rewrite Eq. (2.13) in a form of
$$\eqalign{(xy&e^j)(zue_j)v + (xye^j)(vze_j)u - (xye^j)(uve_j)z\cr
&= \varepsilon\ uv [(xye^j)ze_j] - 3 \lambda \varepsilon\ uv (xyz)\cr
&\quad - \lambda \varepsilon <u \vert z> (xye^j)e_jv +
 \lambda <v \vert z>(xye^j)e_j u\cr
&\quad + \lambda <xye^j \vert zue_j>v - \lambda \varepsilon <xye^j \vert zv
e_j>u + \lambda <xye^j \vert uve_j>z\cr}\eqno(2.16)$$
after some calculations.  Note that both sides of Eq. (2.16) are manifestly
antisymmetric for $\varepsilon =1$ and symmetric for $\varepsilon
 = -1$ with respect to the exchange $u \leftrightarrow v$.

Next, we let $u \rightarrow v \rightarrow z \rightarrow u$ in Eq. (2.16)
and add it to Eq. (2.16) to obtain
$$\eqalign{2(xy&e^j)(zue_j)v \cr
&= \varepsilon\ uv [(xye^j)ze_j] + \varepsilon\ vz [(xye^j)ue_j]\cr
&\quad - 3 \lambda \varepsilon\ uv (xyz)  - 3 \lambda \varepsilon\
 vz(xyu) + \lambda \varepsilon <z \vert v>(xye^j)e_ju\cr
&\quad - \lambda <u \vert v>(xye^j)e_jz + 2 \lambda <xye^j \vert zue_j>v\cr
&\quad + 2 \lambda <xye^j \vert
uve_j>z - 2 \lambda \varepsilon <xye^j \vert zve_j>u \quad .\cr}\eqno(2.17)
$$
Using Eqs. (1.13a)--(1.13c), we can still simplify Eq. (2.17) to be
$$\eqalign{2(xye^j)&(zue_j)v\cr
&= \varepsilon [(xye^j)ze_j]uv - [(xye^j)ue_j]zv - 3 \lambda \varepsilon
(xyz)uv + 3 \lambda (xyu)zv \cr}\eqno(2.18)$$
which is equivalent to the desired relation Eq. (2.11).

The relations in Corollary 1 are necessary to satisfy the identity
$$(xye^j)(zue_j)v = - (zue^j)(xye_j)v$$
which is antisymmetric for $x \leftrightarrow z$ and $y \leftrightarrow
 u$.  These can be derived from Eqs. (1.18) and (2.3e) for example, by
letting $z \leftrightarrow v$ in Eq. (1.18).  We note that the 1st relation
in the Corollary can also be rewritten in a more symmetrical form of
$$\varepsilon [x,y,z]uv = [x,y,u]zv + [x,u,z]yv + [u,y,z]xv \quad .$$

This completes the proof of the Proposition 2. $\ \bull$
\smallskip
\noindent \underbar{Proposition 3}

Suppose that $x \cdot y \cdot z$ is trivial in the sense that it is given
in a form of
$$x \cdot y \cdot z = \gamma \{ <y \vert z>x - <z \vert x>y\} \eqno(2.19)$$
for a constant $\gamma$.  Then, we must have $\gamma =0$ and hence
$x \cdot y \cdot z = 0$ identically, unless we have either $\epsilon
 N = 4$ or $[x,y,z] = 0$ identically.  Especially, the last condition
implies that the original OTS or STS must be trivial.  Conversely, if
$[x,y,z] = 0,\ {\rm then}\ x \cdot y \cdot z$ satisfies Eq. (2.19) with
$\gamma = - {1 \over 3}\ \lambda^2 (\epsilon N-10)$.
\smallskip
\noindent \underbar{Proof}

Since its proof is a bit complicated, we will divide it into the following
three 3 steps by assuming the validity of Eq. (2.19).
\smallskip
\noindent \underbar{Step 1}

For 4-vectors $w,u,v,z\ \epsilon\ V$, and for a basis vector
$e_j$ of $V$, we have
\vfil\eject
\smallskip
$$\eqalignno{w(uv&e^k)(e_je_kz)\cr
&= {1 \over 6}\ \lambda \epsilon (\epsilon N-16)(uvz)e_jw -
{1 \over 6}\ \lambda (\epsilon N-10) w(uve_j)z\cr
&\quad + 2 \lambda \epsilon \ uv(ze_jw) + 2 \lambda \epsilon\ w(uvz)e_j
 + \lambda (wuv)e_jz\cr
&\quad + 2 (\lambda)^2  <v \vert w>ue_jz -  (\lambda)^2
 <u \vert v>we_jz -
 (\lambda)^2 \epsilon<u \vert w>ve_jz\cr
&\quad + {1 \over 6}\ (\lambda)^2 (\epsilon N-16)\{-2 <z \vert w> uve_j +
\varepsilon <uvz \vert e_j>w\}&(2.20)\cr
&\quad + {1 \over 3}\ (\lambda)^2 (\epsilon N-10)<e_j \vert z>uvw  -
2 (\lambda)^2 \epsilon <w \vert uvz>e_j\cr
&\quad + {1 \over 6}\ (\lambda)^2 (\epsilon N-10)<w \vert uve_j>z  +
 \lambda \gamma [<v \vert
z><e_j \vert u>\ - \ <z \vert u><e_j \vert v>]w\cr
&\quad + \lambda \gamma <z \vert e_j>[\varepsilon <v \vert w>u\  - <u \vert w
>v] + {1 \over 2}\ \gamma [\varepsilon <v \vert z>ue_jw \cr
&\quad - \varepsilon <z \vert u>ve_jw - <v \vert e_j>uzw\  +
\ <e_j \vert u>vzw] \quad . \cr}$$
We first calculate
$$e_je_kz = ze_je_k + 2 \lambda <e_k \vert z>e_j
 - \lambda <e_j \vert e_k>z - \lambda <z \vert e_j>e_k$$
to find
$$\eqalign{w(uve^k)(e_je_kz) &= - \varepsilon (uve^k)w(e_je_kz)\cr
&= - \varepsilon (uve^k)w(ze_je_k) -
 2 \lambda (uvz)we_j\cr
&\quad + \lambda \epsilon (uve_j)wz + \varepsilon \lambda
<z \vert e_j>(uve^k)we_k\cr
&= - \varepsilon\{ - \varepsilon (uve^k)(ze_je_k)w +
 2 \lambda <w \vert ze_je_k> uve^k\cr
&\quad - \lambda <uve^k \vert w> ze_je_k - \lambda
<z e_j e_k \vert uve^k>w\}\cr
&\quad - 2 \lambda (uvz)we_j + \lambda \epsilon (uve_j)wz\cr
&\quad + \varepsilon \lambda <z \vert e_j>\{ {1 \over 3}\ \lambda
(\epsilon N-7)uvw + \gamma [<v \vert w>u - \ <w \vert u>v]\}\quad .\cr}$$
For the first term $(uve^k)(ze_je_k)w$, we use the Proposition 1 with the
replacement $x \rightarrow u \rightarrow e_j \rightarrow e_k$ and
$y \rightarrow v \rightarrow w$ in Eq. (2.11).  Also, we note for example
$$<w \vert ze_je_k>uve^k = - uv (z e_j w)$$
\noindent from Eq. (2.1d).
  Then, after some calculations, we obtain Eq. (2.20).
\smallskip
\noindent \underbar{Step 2}

We have the validity of the following relation:
$$\eqalign{(xy&e^j)(uve^k)(e_je_kz)\cr
\noalign{\vskip 4pt}%
&= {1 \over 36}\ \varepsilon (\lambda)^2 \{ (\epsilon N)^2 + 16 \epsilon N
- 224\} \{xy(uvz) + uv (xyz)\}\cr
\noalign{\vskip 4pt}%
&\quad + {1 \over 18}\ (\lambda)^3 (\epsilon N-10)(\epsilon N-16)<v \vert
xyu>z\cr
\noalign{\vskip 4pt}%
&\quad + {1 \over 2}\ \gamma \{ -uz(xyv) + \epsilon vz (xyu)\}\cr
\noalign{\vskip 4pt}%
&\quad - {1 \over 12}\ \gamma \lambda (\epsilon N -16)\{ \epsilon <y \vert
u>xvz - \varepsilon <u \vert x> yvz\cr
\noalign{\vskip 4pt}%
&\qquad - \ <y \vert v> xuz\  + \ <v \vert x> yuz\}\cr
\noalign{\vskip 4pt}%
&\quad + {1 \over 6}\ \epsilon \gamma \lambda (\epsilon N -1)\{<v \vert
z>xyu\  - <z \vert u> xyv\}\cr
\noalign{\vskip 4pt}%
&\quad + 2 \epsilon \gamma \lambda \{ <y \vert z> uvx\  - <z \vert x>
uvy\}\cr
\noalign{\vskip 4pt}%
&\quad + {1 \over 6}\ \epsilon \gamma \lambda (\epsilon N -4)\{<y \vert
uvz>x - \epsilon <x \vert uvz> y\}\cr
\noalign{\vskip 4pt}%
&\quad + \gamma \lambda \{<z \vert
xyu>v\  - <z \vert xyv>u\}\cr
\noalign{\vskip 4pt}%
&\quad + {1 \over 6}\ \gamma (\lambda)^2 (\epsilon N-10) \{ <y \vert u><v
\vert x> \ - \ <u \vert x> <v \vert y> \}z\cr
\noalign{\vskip 4pt}%
&\quad + {1 \over 2}\ \epsilon (\gamma)^2 \{ [<v \vert z><y \vert u> \ -
\ <z \vert u><y \vert v>] x\cr
\noalign{\vskip 4pt}%
&\qquad - [ <v \vert z><u \vert x>\ - \ <z \vert u> <v \vert x>]y\}
\quad .\cr}\eqno(2.21)$$

To prove it, we first set $w = xye^j$ in Eq. (2.20), and we calculate for
example
$$\eqalign{(uvz)e_j (xye^j) &= (xye^j) (uvz)e_j = x \cdot y \cdot (uvz) +
{1 \over 3}\ \lambda (\epsilon N-7)xy(uvz)\cr
\noalign{\vskip 4pt}%
&= {1 \over 3}\ \lambda (\epsilon N-7)xy (uvz) + \gamma
 \{<y \vert uvz>x\  - <uvz \vert x>y \}\cr}$$
\noindent from Eq. (2.1g) with the replacement $z \rightarrow uvz$.
Similarly, we evaluate
$$\eqalign{[(xye^j)uv]e_jz &= - \varepsilon (e_j uv)(xye^j)z
 = (xye^j)(e_juv)z\cr
&= (xye^j)(uve_j)z - 2 \lambda \varepsilon <e_j \vert v>(xye^j)uz\cr
&\quad + \lambda \varepsilon <u \vert e_j>(xye^j)vz +
\lambda \epsilon <v \vert u>(xy e^j)e_jz\cr
&= {1 \over 6}\ \varepsilon \lambda (\epsilon N -16)(xyu)vz -
{1 \over 6}\ \lambda (\epsilon N-16)(xyv)uz\cr
&\quad + {1 \over 2}\ \gamma \epsilon \{<y \vert u>xvz -\ < u \vert
x>yvz\}\cr
&\quad - {1 \over 2}\ \gamma \{<y \vert v>xuz -\ < v \vert
x>yuz\}\cr
&\quad - 2 \lambda (xyv)uz + \lambda \varepsilon (xyu)vz\cr
&\quad - \lambda <v \vert u>\{ {1 \over 3}\ \lambda (\epsilon N-16)xyz
+ \gamma (< y \vert z>x\  - \ <z \vert x>y)\}\cr}$$
\noindent from Eqs. (1.23), (2.2), and (2.11).

Inserting these results, we find Eq. (2.21) after some computations, when
we utilize also Eq. (1.18).
\medskip
\noindent \underbar{Step 3}

In view of Eqs. (1.13b) and (1.21), we note the validity of
$$(xye^j)(uve^k)(e_je_kz) = (uve^j)(xye^k)(e_je_kz)\quad ,\eqno(2.22)$$
which states that it is invariant under $x \leftrightarrow u$ and
$y \leftrightarrow v$.  The consistency of Eq. (2.21) with Eq. (2.22) can
be readily shown to require the validity of
$$\eqalignno{\gamma \varepsilon\{&uv(xyz) - xy(uvz)\}\cr
&= {1 \over 6}\ \gamma \lambda (\epsilon N-16)\{<z \vert v>xyu\  -
 <u \vert z>xyv\  - \ <z \vert y>uvx\  + <x \vert z>uvy\cr
&\quad - <z \vert uvy>x + \varepsilon <z \vert uvx>y\  + <z \vert xyv>u -
\varepsilon <z \vert xyu>v\cr
&\quad - \varepsilon <y \vert u>xvz
 + \epsilon <u \vert x>yvz\  + <y \vert v>xuz\  -
<v \vert x>yuz\} &(2.23)\cr
&\quad + {1 \over 2}\
\epsilon (\gamma)^2
\{[ <v \vert z><y \vert u> - <z \vert u><y \vert
v>]x \cr
&\quad - [<v \vert z><u \vert x> -
<z \vert u><v \vert x>]y
 - [<y \vert z><v \vert x>\cr
&\quad - <z \vert x><v \vert y>]u
 + [<y \vert z><x \vert u> -
<z \vert x><y \vert u>]v\}\quad .\cr}$$
We note especially that all $\gamma$-independent terms have disappeared in
Eq. (2.23).  Therefore, if $\gamma \not= 0$, we must have then
$$\eqalignno{uv(x&yz) - xy(uvz)\cr
&= {1 \over 6}\ \epsilon \lambda (\epsilon N-16)\{<z \vert v>xyu\  -
 <u \vert z>xyv\  - \ <z \vert y>uvx\  + <x \vert z>uvy\cr
&\quad - <z \vert uvy>x + \varepsilon <z \vert uvx>y\  + <z \vert xyv>u -
\varepsilon <z \vert xyu>v\cr
&\quad - \varepsilon <y \vert u>xvz
 + \varepsilon <u \vert x>yvz\  + <y \vert v>xuz\  -
<v \vert x>yuz\} &(2.24)\cr
&\quad + {1 \over 2}\
\gamma
\{[ <v \vert z><y \vert u> - <z \vert u><y \vert
v>]x  - [<v \vert z><u \vert x>\cr
&\quad -
<z \vert u><v \vert x>]y
 - [<y \vert z><v \vert x> - <z \vert x><v \vert y>]u\cr
&\quad + [<y \vert z><x \vert u> -
<z \vert x><y \vert u>]v\}\quad .\cr}$$
Setting $y = e_j$ and $z = e^j$ in  Eq. (2.24), and summing over $j$, it
gives $$(\epsilon N-4) \{ \lambda (\epsilon N-10) uvx - 3 \gamma [<x
\vert u>v\  - <v \vert x>u]\} = 0 \quad . \eqno(2.25)$$
Therefore, we must have either $\epsilon N -4 =0$ or
$$\lambda (\epsilon N-10) uvx - 3 \gamma [<x \vert u>v\  -
 <v \vert x>u] = 0 \quad . \eqno(2.26)$$
Letting $v = e_j$ and $x = e^j$, Eq. (2.26) further leads to
$$(\epsilon N-1) \{ \lambda^2 (\epsilon N-10) + 3 \gamma \}u = 0
\quad .$$
However, since $\epsilon N-1 = 0$ is trivial, this requires
$$\gamma = - {1 \over 3}\ \lambda^2 (\epsilon N-10) \quad . \eqno(2.27)$$
Moreover, if $\gamma \not= 0$, then $\lambda (\epsilon N-10) \not= 0$ so
that Eq. (2.26) together with Eq. (2.27) gives
$$uvx + \lambda [<x \vert u>v\  - <v \vert x>u] = 0$$
\vfil\eject
\noindent or equivalently
 $[u,v,x] = 0$.  Conversely, if we have $[x,y,z]=0$
identically, it is easy to find that $x \cdot y \cdot z$ satisfies
Eq. (2.19) with the value of $\gamma$ being given precisely by Eq. (2.27).
Also, we can verify that the case $\epsilon N = 4$ of the quaternion triple
system which is not however trivial gives $\lambda = 0$ and $\gamma = -2
 \alpha$ in the notation of I.  This completes the proof of the Proposition
3. $\ \bull$
\medskip
\noindent {\bf 3. \underbar{Solution of the Y--B equation}}

Here, we will solve the Y--B  equation (1.8) in a form given by Eq.
(1.25) for either OTS $(\epsilon = +1)$ or STS $(\epsilon = -1)$ under the
additional condition
$$x \cdot y \cdot z = 0 \quad . \eqno(3.1)$$
For simplicity, we write
$$P = P (\theta) \quad , \quad P^\prime = P(\theta^\prime) \quad , \quad
P^{\prime \prime} = P (\theta^{\prime \prime}) \eqno(3.2)$$
and similarly for $Q (\theta),\ R(\theta),\ {\rm and}\ S(\theta)$.
Inserting Eq. (1.25) into Eq. (1.8), each side of Eq. (1.8) is exapnded as
a sum of 64 terms.  The most complicated term is the first one in the
expansion of form
$$P P^\prime P^{\prime \prime} \{ (e_j zu )(xye^j) v -
(e_j xv)(zye^j)u\}$$
which can be, however, simplified by Eq. (2.11) of the Proposition 2
 together with Eqs. (1.13b) and (1.13c).
Utilizing Eq. (1.18)
 and results of the Lemma 1, we find the following expression after
somewhat long calculations:
$$\eqalign{0 &= [v,[u,e_j,z]_{\theta^\prime},
[e^j,x,y]_\theta]_{\theta^{\prime \prime}} - (u \leftrightarrow v,
x \leftrightarrow z, \theta \leftrightarrow \theta^{\prime \prime})\cr
&=  K_1 \{ (yxu)zv - (yzv)xu\} +
K_2 <u \vert v>[x,y,z]
 + K_3 <x \vert z>[y,u,v]\cr
&\quad + K_4 <u \vert x>[y,z,v] -
\hat K_4 <v \vert z> [y,x,u]\cr
&\quad + K_5 <y \vert z>[x,u,v] - \hat K_5 <y \vert x>[z,v,u]\cr
&\quad  + K_6 <u \vert y> [x,z,v] - \hat K_6 <v \vert y> [z,x,u]\cr
&\quad + K_7 <z \vert u>[x,y,v] - \hat K_7 <x \vert v>[z,y,u]\cr
&\quad +K_8 <u \vert [z,y,v]>x - \hat K_8 <v \vert [x,y,u]>z\cr
&\quad + K_9 <v \vert [x,y,z]>u - \hat K_9 <u \vert[x,y,z]>v \cr
&\quad + K_{10} <v \vert y><u \vert z>x - \hat K_{10} <u \vert y> <v \vert
x>z\cr
&\quad + K_{11} <y \vert z><u \vert v>x -  \hat K_{11}
 <y \vert x><v \vert u> z\cr
&\quad  + K_{12} <y \vert z><v \vert x>u -
\hat K_{12} <y  \vert x> <u \vert z> v\cr
&\quad + K_{13} <x \vert z>< y \vert v>u - \hat K_{13} <x \vert z>
<u \vert y>v\cr}\eqno(3.3)$$
where $K_\mu (\mu = 1,2,\dots, 13)$ are cubic polynomials of $P, \
Q,\ R,\ {\rm and}\ S$ to be specified below, and $\hat K_\mu$ is the same
function as $K_\mu$ except for the interchange of $\theta \leftrightarrow
 \theta^{\prime \prime}$.  Note that only $K_1,\ K_2,\ K_3,$ and
$K_{12}$ are self-conjugate, i.e. $\hat K_\mu = K_\mu$ for
$\mu = 1,2,3,\ {\rm and}\ 12$.  The explicit expressions for
$K_\mu$'s are given by
$$\eqalign{K_1 &= - {1 \over 6}\ \lambda (\epsilon N -4) P^{\prime \prime}
 P^\prime P + P^{\prime \prime} R^\prime P - P^{\prime \prime}
P^\prime R - R^{\prime \prime} P^\prime P \quad ,\cr
K_2 &= - {1 \over 3}\ \lambda^2 (\epsilon N-10) P^{\prime \prime} P^\prime
P - \lambda P^{\prime \prime} P^\prime R - \lambda R^{\prime \prime}
P^\prime P -
 2 \lambda P^{\prime \prime} Q^\prime P + P^{\prime \prime} Q^\prime
R + R^{\prime \prime} Q^\prime P \quad ,\cr
K_3 &= - \varepsilon K_2 \quad ,\cr
K_4 &= 0 \quad ,\cr
K_5 &= - {1 \over 6}\ \lambda^2 (\epsilon N-10) P^{\prime \prime} P^\prime
P - {1 \over 3}\ \lambda (\epsilon N-7)
 Q^{\prime \prime} P^\prime P - \lambda P^{\prime \prime}
Q^\prime P -
 \lambda R^{\prime \prime} P^\prime P\cr
&\quad -  Q^{\prime \prime} P^\prime
R + Q^{\prime \prime} P^\prime S - Q^{\prime \prime} Q^\prime P +
Q^{\prime \prime} R^\prime P - P^{\prime \prime}
Q^\prime S  \quad ,\cr
K_6 &= - 2 \lambda  P^{\prime \prime} P^\prime
R + R^{\prime \prime} P^\prime S - P^{\prime \prime}
R^\prime S \quad ,\cr}$$

$$\eqalignno{K_7 &=
 {1 \over 3}\ \lambda^2 (\epsilon N-16) P^{\prime \prime} P^\prime
P + {1 \over 3}\ \lambda (\epsilon N-16)
 P^{\prime \prime} S^\prime P + 2 \lambda P^{\prime \prime}
P^\prime R\cr
&\quad  + 2 \lambda R^{\prime \prime} P^\prime
P + S^{\prime \prime} P^\prime S + P^{\prime \prime}
S^\prime R +
 R^{\prime \prime} S^\prime P -
 P^{\prime \prime} S^\prime S - S^{\prime \prime} S^\prime P
 \quad ,\cr
K_8 &= \varepsilon K_6 \quad ,&(3.4)\cr
K_9 &= - \hat K_5 \quad , \cr
K_{10} &= {1 \over 3}\ \lambda^3 (\epsilon N-16) P^{\prime \prime} P^\prime
P + {1 \over 3}\ \lambda^2 (\epsilon N-16)
 P^{\prime \prime} S^\prime P + \lambda (P^{\prime \prime}
S^\prime + S^{\prime \prime} P^\prime)R\cr
&\quad - \lambda (S^{\prime \prime} R^\prime
 - R^{\prime \prime} S^\prime + S^{\prime \prime}
S^\prime)P +  \lambda (S^{\prime \prime}
P^\prime - P^{\prime \prime} S^\prime \cr
&\quad + R^{\prime \prime} P^\prime - P^{\prime \prime} R^\prime )S +
R^{\prime \prime}S^\prime S + S^{\prime \prime} S^\prime R -
S^{\prime \prime} R^\prime S\cr
K_{11} &= -{1 \over 6}\ \lambda^3 (\epsilon N-10) P^{\prime \prime} P^\prime
P + {1 \over 3}\ \lambda^2 (\epsilon N-7)
 Q^{\prime \prime} P^\prime P - \lambda^2 P^{\prime \prime}
Q^\prime P\cr
&\quad + \lambda (Q^{\prime \prime} Q^\prime
 + R^{\prime \prime} Q^\prime - Q^{\prime \prime}
R^\prime)P +  \lambda (Q^{\prime \prime}
P^\prime - P^{\prime \prime} Q^\prime)S \cr
&\quad + R^{\prime \prime} Q^\prime S - Q^{\prime \prime} R^\prime S -
Q^{\prime \prime}Q^\prime R\quad ,\cr
K_{12} &= {1 \over 6}\ \lambda^3 (\epsilon N-10) P^{\prime \prime} P^\prime
P + \lambda^2
 P^{\prime \prime} Q^\prime P - \lambda (\epsilon N-1) Q^{\prime \prime}
P^\prime Q - \lambda (S^{\prime \prime} P^\prime +
R^{\prime \prime} P^\prime)Q\cr
&\quad + \lambda (P^{\prime \prime} Q^\prime
 - Q^{\prime \prime} P^\prime)S  + \lambda S^{\prime \prime}
Q^\prime P -  \lambda Q^{\prime \prime}
P^\prime R - (\epsilon N) Q^{\prime \prime} S^\prime Q \cr
&\quad - (Q^{\prime \prime} Q^\prime + Q^{\prime \prime} R^\prime +
R^{\prime \prime} S^\prime
 + S^{\prime \prime} S^\prime )Q - (Q^{\prime \prime} S^\prime -
S^{\prime \prime} Q^\prime ) S - Q^{\prime \prime} S^\prime R \quad ,\cr
K_{13} &= \hat K_{11} \quad .\cr}$$

We note that if we had had used
$$x \cdot y \cdot z = \gamma \{ <y \vert z >x\  - < z \vert x>y\}$$
instead of Eq. (3.1), then $K_4$, for example, would become
$$K_4 = - {1 \over 2}\
\epsilon \gamma P^{\prime \prime} P^\prime P$$
instead of zero as in Eq. (3.4).  Simlarly, we must add extra term
$${1 \over 2}\ \gamma P^{\prime \prime} P^\prime P$$
to $K_6$, but \underbar{not} to $K_8$, so that the relation $K_8 =
\varepsilon K_6$ will not hold any longer.
 For the octonionic triple product, we can further reduce the first
 term proportional to $K_1$ by using the identity given in I.
  Then, we can verify after some calculations that Eqs.
(3.3) and (3.4) of the present paper reproduce the corresponding equations in
I for $\varepsilon N = 8$, $\lambda = -3 \beta$, with $\alpha = \beta^2$.

At any rate, the Yang--Baxter equation will be automatically satisfied,
 if we have eight equations
$$K_1 = K_2 = K_5 = K_6 = K_7 = K_{10} = K_{11} = K_{12} = 0
\quad .\eqno(3.5)$$
First, consider $K_1 = 0$ which can be rewritten as
$$R^\prime / P^\prime - R/P - R^{\prime \prime} / P^{\prime \prime} =
 {1 \over 6}\ \lambda (\epsilon N-4) \quad .$$
However, since $\theta^\prime = \theta + \theta^{\prime \prime}$,
 this equation requires the validity of
$$R (\theta)/ P(\theta) = a + b \theta \eqno(3.6)$$
where we have set
$$a = - {1 \over 6}\ \lambda (\epsilon N - 4) \eqno(3.7)$$
and $b$ is an arbitrary constant.
Then, $K_2 = 0$ which can be rewritten as
$$- {1 \over 3}\ \lambda^2 (\epsilon N -10) -
\lambda \left( {R \over P} + {R^{\prime \prime} \over P^{\prime \prime}}
\right) + {Q^\prime \over P^\prime} \
\left( {R^{\prime \prime} \over P^{\prime \prime}} + {R \over P}
-  2 \lambda \right) = 0$$
can be solved to yield
$$Q(\theta) /P(\theta) = \lambda -
{2 a \lambda \over 2(a - \lambda) + b \theta}\quad . \eqno(3.8)$$
Similarly, the condition $K_6 = 0$ which is equivalent to
$$- 2 \lambda \ {R \over P} + \left( {R^{\prime \prime} \over
P^{\prime \prime}} - {R^\prime \over P^\prime}
\right) \ {S \over P} = 0$$
leads to
$$S(\theta) /P(\theta) = - 2 \lambda - {2
\lambda a \over b \theta} \quad . \eqno(3.9)$$
These determine $R(\theta),\ Q(\theta),\ {\rm and}\ S(\theta)$ in terms of
$P(\theta)$.  The rest of relations $K_5 = K_7 = K_{10} = K_{11}
 = K_{12} = 0$ can be then verified to be automatically satisfied after
some computations.   Rewriting
$$\eqalign{[z,x,y]_\theta = &P(\theta) [x,y,z] + A(\theta) <x \vert
y>z\cr
&+ B(\theta) <z \vert x>y + C(\theta) <z \vert y>x \quad , \cr}
\eqno(3.10)$$
then we find
$$\eqalignno{A(\theta) &= \big\{ \lambda - {2a \lambda \over 2( a -
\lambda) + b \theta}\big\} \ P(\theta) \quad , &(3.11a)\cr
\noalign{\vskip 4pt}%
B(\theta) &= \big\{ (a-\lambda) + b \theta \big\}\ P(\theta)
\quad , &(3.11b)\cr
\noalign{\vskip 4pt}%
C(\theta) &= \big\{ - \lambda - {2 a \lambda \over
b \theta}\big\} \ P(\theta) \quad , &(3.11c)\cr}$$
because of Eqs. (1.28).  This reproduces Eq. (1.29).  As we will see in the
next section, the case of the octonionic triple product corresponding to
$\varepsilon N=8$ with the normalization $\lambda = -3 \beta = 3$ as in I
satisfies the desired condition $x \cdot y \cdot z = 0$.  In that case,
$a = -2$ by Eq. (3.7) and hence
$$\eqalignno{A(\theta)/P(\theta) &= {18 - 3 b \theta \over
10 - b \theta} \quad , &(3.12a)\cr
\noalign{\vskip 4pt}%
B(\theta)/P(\theta) &= b \theta - 5 \quad , &(3.12b)\cr
\noalign{\vskip 4pt}%
C(\theta)/P(\theta) &= {12 - 3 b \theta \over
b \theta} \quad ,&(3.12c)\cr}$$
which reproduces the result of I as well as that given by de Vega and
Nicolai$^{13)}$.  This fact serves as a cross--check of our calculations,
since these latter computations are based upon entirely different method.

In our derivation of Eq. (3.11), we have implicitly assumed $P(\theta)
\not= 0$.  However, if we have $P(\theta) = 0$ identically, the situation
becomes simpler, since we need then consider only 3 conditions
$K_{10} = K_{11} = K_{12} = 0$.  Assuming $C(\theta) \not= 0$, the solution
is given now by
$$[z,x,y]_\theta = A(\theta) <x \vert y>z + B(\theta) <z \vert x>y +
C(\theta)<z \vert y>x \quad , \eqno(3.13a)$$
with
$${B (\theta) \over C(\theta)} = b \theta \quad , \quad
{A (\theta) \over C(\theta)} = - {2 b \theta
 \over 2 b \theta + (\epsilon N -2)} \eqno(3.13b)$$
for arbitrary constant $b$.  Note that the dimension $N$ is completely
arbitrary, and that since $P(\theta) = 0$,
 we need no longer assume here
that $V$ is either OTS or STS.  Also, Eq. (3.13b) reproduces the result Eq.
(4.12) of I for $\varepsilon N = 4$.  In this case, we can also forget
about the condition $x \cdot y \cdot z = 0$. We remark that Eq. (3.13b) for
$\varepsilon = 1$ reproduces the result of Zamolodchikov's so(N)
model$^{14)}$, while $\varepsilon = -1$ corresponds to sp(N) symmetry.

We may interpret Eq. (3.13a) as the condition $[x,y,z] = 0$ rather than
$P(\theta) = 0$.  Then, the condition $x \cdot y \cdot z = 0$ can be
achieved only if $\epsilon N = 10$ by the Proposition 3.  The solution Eqs.
(3.11) for $\varepsilon N = 10$ agree, of course, with Eqs. (3.13), when
we suitably renormalize $\lambda \ {\rm and}\ b$, for example by
$a = - \lambda = 1$.

Also if we have $C(\theta) = 0$ in Eq. (3.13a), then the solution of the
Y--B equation is rather trivial with $A(\theta) = 0$ and $B(\theta)$ being
 arbitrary.

Finally, the case $N = 2$ for STS is special,
 and we can find a more general trigonometric solution in that case, as we
will explain in section 5.

\medskip
\noindent {\bf 4. \underbar{Condition $x \cdot y \cdot z = 0$}}

In this section, we seek OTS or STS satisfying the condition
$x \cdot y \cdot z = 0$.  For the trivial case of $[x,y,z] = 0$
identically, it is, of course, satisfied only for $\varepsilon N = 10$
 (i.e. $\varepsilon = 1 \ {\rm and}\ N = 10$) by the Proposition 3.  Also
for the octonionic triple product corresponding to $\varepsilon = 1$ and
$N=8$, we can verify $x \cdot y \cdot z = 0$ by using the result of I.
Similarly, for simple cases of $N=2$ and $N=4$ for STS
 to be given shortly, we can directly
show the same by explicit computations.  However, the task will become
increasingly unmangeable for larger values of $N$.

We can nevertheless give a simple characterization of OTS or STS satisfying
$x \cdot y \cdot z = 0$ as follows.  For this end, we utilize the method
explained in ref. 15 for STS and also briefly in I for OTS.  For many STS
and OTS, the underlying vector space $V$ is often a module of a Lie algebra
$L$, which we will assume in this section.  Also, unless we state
otherwise, we assume $L$ to be simple and $V$ to be an irreducible module
of $L$.  Let $W_1$ and $W_2$ be two $L$-modules which need not be, however,
irreducible.  We denote then by Hom$(W_1 \rightarrow W_2)$ be the vector
space of all homomorphism from $W_1$ to $W_2$, which are compatible with
the action of $L$.

Next, the tensor product $V \otimes V \otimes V$ can be decomposed into a
sum of vector spaces with distinct permutation symmetries and we set as in I
$$V = [1] \quad , \quad (V \otimes V \otimes V)_S = [3] \quad , \quad
(V \otimes V \otimes V)_A = [1^3] \quad , \quad
(V \otimes V \otimes V)_M = [2,1] \eqno(4.1)$$
etc, where the suffices $S,\ A,\ {\rm and}\ M$ refer to the totally
symmetric, antisymmetric and mixed symmetries with respect to the
permutation group $Z_3$, and the symbol $[f_1,f_2,f_3]$ with
 $f_1 \geq f_2 \geq f_3 \geq 0$ designates the standard Young-tableau
notation$^{16)}$.  Suppose that we have
$${\rm Dim\ Hom}\ ([1^3] \rightarrow [1]) = 1 \eqno(4.2)$$
or
$${\rm Dim\ Hom}\ ([3] \rightarrow [1]) = 1 \quad , \eqno(4.3)$$
where Dim $W$ is the dimension of vector space $W$.  This implies$^{15)}$,
then, the existence of unique $L$--covariant triple product $[x,y,z]$ in
$V$ which is totally antisymmetric for the former, or totally symmetric for
the latter.  Moreover, these products can be shown to satisfy the axioms of
OTS or STS, if some additional conditions such as
$${\rm Dim\ Hom}\ ([4,1] \rightarrow [1]) \leq 2$$
etc. hold valid.  However, since these are discussed in detail in ref. 15
and also in I, we will not go into detail.

Returning now to the dotted product $x \cdot y \cdot z$, the validity of
Eqs. (2.3a) and (2.3b) implies that $x \cdot y \cdot z$ is
 contrarily an element of
$${\rm Hom}\ ([2,1] \rightarrow [1]) \quad .$$
Suppose that we have
$${\rm Dim\ Hom}\ ([2,1] \rightarrow [1]) \leq 1 \eqno(4.4)$$
in addition.  Then, following the reasoning of ref. 15, $x \cdot
y \cdot z$ can be rewritten in the form of Eq. (2.19) for some constant
 $\gamma$, since $<y \vert z>x\  - \ <z \vert x>y$ can be easily seen also to
be an element of Hom$([2,1] \rightarrow [1])$.  In that case, the
Proposition 3 guarantees $x \cdot y  \cdot z = 0$ identically, provided
that the original OTS or STS is not trivial, i.e. $[x,y,z] \not= 0$ with
 $\varepsilon N \not= 4$.  Therefore, we have only to verify the validity of
Eq. (4.4).  For OTS, both octonionic $(N=8)$ and Malcev $(N=7)$ triple
products have been shown in I to satisfy the condition (4.4) so that we
have $x \cdot y \cdot z = 0$ identically.  As we have already stated, this
can be easily verified also by a direct computation for the former case of
$N=8$.  Therefore, two cases of $N=7$ and 8 with $\varepsilon = 1$ furnish
solutions of the Y--B equation.  Although there may exist other OTS
 satisfying the condition, we do
not know yet.  Also the case of $N=4$ corresponding to the quaternionic
triple product does not satisfy $x \cdot y \cdot z = 0$.  However, we have
already found the solution for this case in I by a different method.

We will devote the rest of this section to the STS case of $\varepsilon
 = -1$.  It is now known that there exists intimate inter--relationship
between a simple Lie algebra other than $A_1$ and a STS.  Especially,
Asano$^{17)}$ shows that we can construct any simple Lie algebra $L_0$ from
some STS, and conversely that a STS can always be constructed from any
simple Lie algebras $L_0$
other than $A_1$.  Therefore, we can construct STS's from any simple Lie
algebras, following the method of the ref. 17.  However, as we will show
shortly, only STS's constructed from $L_0 = A_2,\ G_2,\ F_4,\ E_6,\ E_7,\
{\rm and}\ E_8$ satisfy the desired condtion $x \cdot y \cdot z = 0$.  Let
$H_0$ and $\alpha\  \epsilon\  \Delta$ be respectively
 the Cartan sub-algebra in Chevalley
basis and non-zero root of a complex simple Lie algebra $L_0$, where
$\Delta$ is the root-system with respect to some lexicographical
ordering$^{18)}$.  Let $\rho$ be the highest root normalized to
$$(\rho , \rho) = 2 \quad . \eqno(4.5)$$
Setting
$$h = [E_\rho , E_{- \rho}]\quad, \eqno(4.6)$$
we can decompose $L_0$ into a direct sum
$$L_0 = V_2 \oplus V_1 \oplus V_0 \oplus V_{-1} \oplus V_{-2} \eqno(4.7)$$
which satisfies moreover
$$[h,x_n] = n x_n \quad , \quad  {\rm if}\quad x_n \  \epsilon
 \  V_n
 \quad , \eqno(4.8a)$$
and
$$[V_n, V_m] \subset V_{n+m} \eqno(4.8b)$$
for $n,m = 0, \ \pm 1, \ {\rm and}\ \pm 2$.  More explicitly,
$$\eqalign{V_2 &= \{x \vert x = c E_\rho \}\quad , \quad
V_{-2} = \{x \vert x = c E_{- \rho} \} \quad ,\cr
V_1 &= \{ x \vert x = \sum_\alpha c_\alpha E_\alpha ,
(\rho , \alpha ) = 1\} \quad , \cr
V_{-1} &= \{ x \vert x = \sum_\alpha c_\alpha E_\alpha, (\rho , \alpha)
= -1 \} \quad ,\cr
V_0 &= \{ x \vert x = x_0 + \overline x \quad , \quad
x_0 \epsilon  H_0 \quad , \quad \overline x = \sum_\alpha c_\alpha E_\alpha
 \ {\rm with}\ (\rho , \alpha ) = 0\}\cr}\eqno(4.9)$$
for constants $c$, and $c_\alpha$'s.

We identify our module $V$ to be
$$V = V_1 \eqno(4.10)$$
by a reason to be given shortly.  First for any $x,\ y\  \epsilon
\  V_1$, there
exists an inner product $< x \vert y>$ defined by
$$[x,y] = 2 < x \vert y > E_\rho \eqno(4.11)$$
since the left side must belong to the space $V_2$ by Eq. (4.8b).  Clearly,
$< x \vert y>$ is non-degenerate in $V_1$ and satisfies the symplectic
condition
$$< x \vert y> \ = -<y
\vert x> \quad . \eqno(4.12)$$
Following Asano, we then introduce a triple product $xyz$ in $V = V_1$ by
$$xyz = {1 \over 2}\ \left\{ [z,[y,[x, E_{- \rho}]]] +
[z,[x,[y, E_{- \rho}]]]\right\} \quad . \eqno(4.13)$$
The fact that $xyz \ \epsilon\ V_1$ follows again from Eq. (4.8b).  It is
easy to show the validity of
$$\eqalignno{xyz &= yxz &(4.14a)\cr
xyz - xzy &= 2< y \vert z> x\  - <x \vert y>z\  -
 \ < z \vert x>y \quad . &(4.14b) \cr}$$
Finally, Asano$^{17)}$ also proves  the derivation relation
$$uv(xyz) = (uvx)yz + x(uvy)z + xy(uvz) \eqno(4.15)$$
for the product.  Therefore, the product $xyz$ defines a symplectic triple
system with $\lambda$ normalized to be 1.

Actually, $V = V_1$ can be regarded as a module of a sub-Lie algebra $L$ of
$L_0$.  In general, $L$ is a reductive Lie algebra, i.e., it is a direct
sum of semi-simple and Abelian algebras.  However, for our purpose, we
regard it as a simple Lie algebra by choosing only its maximal simple
algebra and deleting all others contained therein.
 The Dynkin diagram of $L$ is precisely the one obtained from the general
Dynkin diagram of $L_0$, by deleting simple roots connected with the
lowest root $- \rho$.  We are now in a position to characterize $L,\ V,$
and $N = \ {\rm Dim}\ V$ for any $L_0$.  First for any classical Lie
algebras $L_0 = A_n (n \geq 3),\ B_n (n \geq 2),\ C_n (n \geq 2)$ and
$D_n (n \geq 3)$, we find respectively
$$L = A_{n-2}, B_{n-1}, C_{n-1},\ {\rm and}\ D_{n-1} \quad .$$
However the condition $x \cdot y \cdot z = 0$ is not satisfied by any of
these by the following reason.  For $L = C_{n-1}$, $V$ is a $N = 2(n -1)$
dimensional irreducible module of $L$.  However, the resulting STS turns
out to be trivial.  On the other sides, we find that $V$'s for
$L = A_{n-2}$, $B_{n-1}$, and $D_{n-1}$ are reducible and that the
condition Eq. (4.4) is not satisfied.  The exception is for  $L_0 = A_2$
 where $L$ is null with $N=2$.  In this case, we can verify
 $x \cdot y \cdot z = 0$ by a direct computation.  Next, consider 5
exceptional Lie algebras $L_0 = G_2, \ F_4,\ E_6,\ E_7,\
{\rm and}\ E_8$ where we find
\item{(i)} $L_0 = G_2\quad , \quad L = A_1 \quad , \quad
 N=4\quad ,$
\item{(ii)} $L_0 = F_4 \quad , \quad L = C_3 \quad ,
\quad N=14 \quad ,$
\item{(iii)} $L_0 = E_6 \quad , \quad L = A_5 \quad ,
\quad N = 20 \quad ,$ \hfill (4.16)
\item{(iv)} $L_0 = E_7 \quad , \quad L = D_6 \quad , \quad N = 32
\quad ,$
\item{(v)} $L_0 = E_8 \quad , \quad L = E_7 \quad , \quad N = 56 \quad .$

\noindent Moreover, $V$ for all
 these cases are found to be irreducible $L$-modules,
satisfying the desired  condition
$${\rm Dim \ Hom}\ ([2,1] \rightarrow [1]) = 1$$
so that we have $x \cdot y \cdot z = 0$ automatically for all these cases.
We have also explicitly verified it for the simplest case of $N=4$ for $L =
A_1$. The fact that 5 cases listed in Eq. (4.16) defines indeed STS's can
be also directly shown by the method given in ref. 15.

In conclusion, 6 cases of $N=2,\ 4,\ 14,\ 20,\ 32,\ {\rm and}\ 56$ with
$\varepsilon = -1$ furnish solutions of the Y--B equation.  In this
connection, we note that both irreducible module $V$ for $L = A_5$ and
$C_3$ in Eq. (4.16)
correspond to totally antisymmetric Young tableau $[1^3]$ of su(6)
and sp(6) with respect to their 6--dimensional fundametal representations,
while the 32--dimensional module for $L = D_6$ is its basic spinor
representation.  The case of $L = E_7$ with $N=56$ refers, of course, to
the Freudenthal's triple system.

Finally, $N=4$ for $L=A_1$ is realizable as the 4--dimensional spin 3/2
representation of the so(3) $\simeq$ su(2) algebra.  Let $x_M$ with
$M = 2m = 3,\ 1,\ -1,\ -3$ be its basis, corresponding to the magnetic
quantum number $m = 3/2,\ 1/2,\ -1/2,\ {\rm and}\ -3/2$, respectively.  We
may normalize them according to
$$<x_M \vert x_{M^\prime}>\ = - {M \over 2}\
\delta_{M + M^\prime, 0}\quad . \eqno(4.17)$$
Then, the totally symmetric triple product $[x,y,z]$ satisfies
$$[x_{M_1},x_{M_2},x_{M_3}] =
 C(M_1, M_2, M_3) x_{M_1 +M_2 +M_3} \eqno(4.18)$$
for a constant $C(M_1,M_2,M_3)$ where we set $x_M = 0$ identically, unless
we have $M \not= 3,\ 1,\ -1,\ {\rm or}\ -3$.  The physical meaning of
$C_{M_1,M_2,M_3}$ is that it is precisely the Clebsch--Gordan coefficient
of totally symmetric tensor product $(V \otimes V \otimes V)_S$ into the
unique spin 3/2 representation $V = \{3/2\}$.  Indeed, we calculate
$$(V \otimes V \otimes V)_S = \{9/2\}
\oplus \{5/2\} \oplus \{3/2\}$$
where $\{j\}$ designates the irrreducible module of su(2) with spin
$j = 0,\ {1 \over 2},\ 1,\ {3 \over 2},\ 2,\ \dots$.  Non-zero
Clebsch--Gordan coefficients $C(M_1,M_2,M_3)$ are tabulated below to be
$$\eqalign{C(1,1,1) &= C(-1,-1,-1) = {2 \over 3}\quad ,\cr
C(-1,-1,1) &= -C(-1,1,1) = {1 \over 3} \quad ,\cr
C(-1,-1,3) &= C(1,1,-3) = -2 \quad ,\cr
C(-1,1,-3) &= C(-1,-3,3) = - {1 \over 2}\quad ,\cr
C(-1,1,3) &= C(1,-3,3) = {1 \over 2} \quad ,\cr
C(-3,3,3) &= -C(-3,-3,3) = 3 \quad .\cr}\eqno(4.19)$$
All other cases except for permutations of $M_1,\ M_2,\ {\rm and}\
M_3$ in the above list give zero value for $C(M_1 , M_2, M_3)$.  We note
that $C(M_1, M_2, M_3)$ is totally symmetric in $M_1,\ M_2,\ {\rm and}
\ M_3$.  Moreover it admits an automorphism $\sigma\ :\ V \rightarrow
V$
$$\eqalignno{\sigma (x_M) &= \varepsilon (M) x_{-M} \quad ,
&(4.20a)\cr
\varepsilon (M) &= \cases{1 \quad ,\quad ${\rm for}\quad  $M > 0$$\cr
\noalign{\vskip 4pt}%
-1 \quad ,\quad ${\rm for}\quad  $M < 0$$\cr}&(4.20b)\cr}$$
which satisfies
$$\eqalignno{&\sigma ([x,y,z]) = [\sigma (x), \sigma (y), \sigma
(z)] &(4.21a)\cr
&<\sigma (x) \vert \sigma (y)>\ = \ <x \vert y> &(4.21b)\cr
&\sigma^2 = - I \quad . &(4.21c)\cr}$$

The $N=4$ triple product possesses $N=2$ sub-algebra consisting of two element
$x_3$ and $x_{-3}$.  One interesting aspect of the STS
 for $N  = 2$ is that the
 following special identity holds valid for the normalization condition
 $\lambda = 1$ with $< x_{-3} \vert x_3 > \ = 3/2$:
$$\eqalign{[x,y,&[u,v,z]] - [u,v,[x,y,z]]\cr
&= \ <x \vert u>[v,y,z]\  + <y \vert v>[u,z,x]\  +
\ <x \vert v> [u,y,z]\cr
&\quad + <y \vert u>[v,z,x] + [<y \vert u><z \vert v>\ + \ <y \vert v><z
\vert u>]x \cr
&\quad + [<z \vert u><x \vert v>\ + \ <z \vert v><x \vert u>]y +
 [<x \vert v><z \vert y>\cr
&\quad + \ <y \vert v><z \vert x>]u + [<z \vert x><y \vert u> \
+\ <z \vert y><x \vert u>]v\cr}\eqno(4.22)$$
as we may verify easily.  Because of Eq. (4.22), we can reduce
$(yxu)zv - (yzv)xu$ in terms of simpler expressions.  Together with other
identities for $\varepsilon N=-2$, this enables for us to considerably
simplify Eq. (3.3) further, and we can find a more general solution for the
Y--B equation.  This will be explained in detail in the next section.
\medskip
\noindent {\bf 5. \underbar{Solution of the STS for N = 2}}

For the special case of $N=2$ for STS, we can find a more general solution
than the one given by Eqs. (1.29).  The reason is first because we have the
special relation Eq. (4.22) for that case.  Second, we note the validity of
the identity
$$<y \vert z>x\  + <z \vert x> y\  + \ <x \vert y>z = 0 \eqno(5.1)$$
for $N=2$ with $\varepsilon = -1$
 since the left side of Eq. (5.1) is totally antisymmetric in
 $x,\ y,\ {\rm and}\ z$.
  Especially, $[z,x,y]_\theta$ given by
 Eq. (1.25) is invariant under the transformation
$$Q(\theta) \rightarrow Q(\theta) + F(\theta) \quad , \quad
R(\theta) \rightarrow R(\theta) + F(\theta) \quad , \quad
Q(\theta) \rightarrow Q(\theta) + F(\theta) \quad , \eqno(5.2)$$
for an arbitrary function $F = F(\theta)$.  However the expression
Eq. (3.3) is not manifestly invariant under it, implying that we may reduce
the expression further into a simpler form.

Utilizing Eq. (5.1), we can moreover note
$$\eqalignno{&<y \vert z><u \vert v>\  = -<y \vert u><v \vert z>\ -
\ <y \vert v><z \vert u> \quad , &(5.3a)\cr
&<u \vert [x,y,z]>v\  - <v \vert [x,y,z]>u =
\ <u \vert v>[x,y,z]\quad , &(5.3b)\cr
&<u \vert v>[x,y,z] = -< v \vert y>[x,u,z]\  -\
<y \vert u>[x,v,z]
&(5.3c)\cr}$$
and so on.  Using these relations as well as Eq. (4.22), we can, now,
simplify the right side of Eq. (3.3) as
$$\eqalign{0 &= [v,[u,e_j,z]_{\theta^\prime},
[e^j,x,y]_\theta]_{\theta^{\prime \prime}} -
(u \leftrightarrow v, x \leftrightarrow z,
\theta \leftrightarrow \theta^{\prime \prime})\cr
&= W_1 <v \vert y>[u,x,z] - \hat W_1 <u \vert y>[v,z,x]\cr
&\quad + W_2 <x \vert y>[u,z,v] - \hat W_2 <z \vert y> [v,x,u]\cr
&\quad + W_3 \{ < u \vert [z,y,v]>x\  +
<v \vert [x,y,u]>z \cr
&\quad + \ <v \vert [x,y,z]>u\  + \ <u \vert [x,y,z]>v\}\cr
&\quad + W_4 <v \vert y><u \vert z>x - \hat W_4 <u \vert y><v \vert x>z\cr
&\quad + W_5 <y \vert u><v \vert z>x - \hat W_5 <y \vert v><u \vert x>z\cr
&\quad  + W_6 <v \vert x><z \vert u>y \cr}\eqno(5.4)$$
where we find
$$\eqalign{W_1 &= 3K_1 - K_2 - \hat K_4 - \hat K_6 + \hat
K_7 + {1 \over 4}\ (K_8 - \hat K_8) + {1 \over 4}\ ( K_9 +
3 \hat K_9)\quad ,\cr
W_2 &= -3K_1 + K_3 -  K_4 + \hat K_5 - \hat
K_7 - {1 \over 4}\ (3K_8 + \hat K_8) + {1 \over 4}\ ( K_9 -
 \hat K_9)\quad ,\cr
W_3 &= {1 \over 4}\ (K_8 - \hat K_8 + K_9 - \hat K_9)\quad ,\cr
W_4 &= 3 K_1 + K_{10} - K_{11} +
\hat K_{12} - K_{13}\quad , \cr
W_5 &= - K_{11} + \hat K_{13}\quad ,\cr
W_6 &= -K_{12} + \hat K_{12} \quad .\cr}\eqno(5.5)$$
However, in view of relations such as $K_{11} = \hat K_{13},\ K_4 = 0,
\ K_3 = K_2,\ K_8 = - K_6,\ {\rm and}\ K_9 = - \hat K_5$ as well as
$\hat K_{12} = K_{12}$ by Eq. (3.4), we have identities
$$W_5 = W_6 = 0 \quad , \quad W_1 + W_2 + 2 W_3 = 0 \quad , \eqno(5.6)$$
so that the Y--B equation is satisfied, provided that  we have 3
equations;
$$W_1 = W_3 =
W_4 = 0 \quad . \eqno(5.7)$$
Moreover, we can verify the fact that $W_1,\ W_3,\ {\rm and}\ W_4$ are
invariant under the transformation Eq. (5.2).  Therefore, choosing
$F= -Q$, we can effectively set $Q = 0$ so that
$$[z,x,y]_\theta = P(\theta)xyz + R(\theta)<z \vert x>y + S(\theta)
 <y \vert z>x \quad .\eqno(5.8)$$
First consider the relation $W_3 = 0$ which can be rewritten as
$$(3 + S/P)(R^\prime / P^\prime - R^{\prime \prime} / P^{\prime \prime}) =
(3 + S^{\prime \prime} / P^{\prime \prime})(R^\prime /
P^\prime - R / P) \quad , \eqno(5.9)$$
assuming $P(\theta) \not= 0$ with $\lambda = 1$.  Then, $W_1 = 0$ together
with $W_3 = 0$ leads similarly to the validity of
$$\eqalignno{(3 &+ S/P)(R^\prime / P^\prime - R^{\prime \prime}
 / P^{\prime \prime}) +
(3 + S^\prime / P^\prime)(R / P +
R^{\prime \prime} / P^{\prime \prime})&(5.10)\cr
&+  (3 + S^{\prime \prime} /P^{\prime \prime})(3 + S/P)
- (3 + S^\prime / P^\prime)(3 + S/P)  - (3 + S^\prime /P^\prime )(3 +
S^{\prime \prime} / P^{\prime \prime}) = 0\cr}$$
while $W_4 = 0$ is rewritten as
$$\eqalign{-9 P^{\prime \prime} P^\prime P &- 6 P^{\prime \prime}
S^\prime P + 3(P^{\prime \prime} R^\prime P - P^{\prime \prime}
P^\prime R - R^{\prime \prime} P^\prime P)\cr
&+ P^{\prime \prime} S^\prime R + S^{\prime \prime} P^\prime
R - S^{\prime \prime} R^\prime P + R^{\prime \prime} S^\prime P -
S^{\prime \prime} S^\prime P + S^{\prime \prime} P^\prime S -
P^{\prime \prime} S^\prime S\cr
&+ R^{\prime \prime} P^\prime S - P^{\prime \prime}
 R^\prime S + R^{\prime \prime} S^\prime S + S^{\prime \prime}
S^\prime R - S^{\prime \prime} R^\prime S = 0\quad .\cr}\eqno(5.11)$$
There exist two distinct classes of solutions.  First, we note that Eqs.
(5.9), (5.10), and (5.11) admit a solution where we have
$$S(\theta) = - 3 P(\theta) \eqno(5.12)$$
while $R(\theta)$ remains arbitrary.  Then, noting
$$xyz = [x,y,z]\  + \ <y \vert z>x\  - \ <z \vert x>y \quad ,$$
this gives the first solution:
$$[z,x,y]_\theta = P(\theta) \{[x,y,z] -
2<y \vert z> x \} + T(\theta)
<z \vert x> y \quad , \eqno(5.13)$$
where both $P(\theta)$ and $T(\theta)$ are arbitrary functions of
$\theta$.  The old solution Eqs. (1.29) with $\lambda = a = 1$ corresponds
to a special choice of
$$T(\theta) = \left( -1 + b \theta + {2 \over
b \theta}\right) P(\theta) \quad .$$
However, since $T(\theta)$ is now arbitrary, we can set $T(\theta) =
 0$ in Eq. (5.13), if we wish.

The second solution is, in constrast, of trigonometric type with
$$\eqalign{R (\theta)/P(\theta) &= 3 \quad ,\cr
S(\theta)/P(\theta) &= -3 + {6 \over 1 - \exp (k \theta)}\cr} \eqno(5.14)$$
where $k$ is an arbitrary constant.  Then, the solution is given by
$$\eqalign{[z,x,y]_\theta = &P(\theta) \{ [x,y,z] +
 2<z \vert x>y\cr
 &+ \big( -2 + {6 \over 1 - \exp (k \theta)}\big) \
<y \vert z>x\} \quad .\cr} \eqno(5.15)$$
Another constant solution can be obtained from this also by letting
 $k \rightarrow - \infty$ for $\theta > 0$ to give
$$[z,x,y]_\theta = P (\theta) \{ [x,y,z] +
2<z \vert x>y + 4 < y \vert z> x \} \quad . \eqno(5.16)$$
We can directly verify that this is also a solution.  Similarly, for
$k \rightarrow + \infty$, Eq. (5.15) will give a special case of
Eq. (5.13).

Note that we cannot here change
 arbitrarily the normalizations of triple products and
inner product since they must satisfy the condition Eq. (4.22).
\medskip
\noindent {\bf 6. \underbar{Concluding Remarks}}

In this paper as well as in the preceeding one, we have found several
solutions of the Yang--Baxter equation in a triple product form under the
ansatz Eq. (1.9).  First, we note that we can dispense with the symmetry
condition now by introducing the $\theta$--dependent triple products
$[x,y,z]_\theta$ and its conjugate $[x,y,z]^*_\theta$ by
$$R^{cd}_{ab} (\theta) = \ <e^c \vert [e^d, e_a, e_b]_\theta >\ = \
<e^d \vert [e^c, e_b,e_a]^*_\theta >\quad . \eqno(6.1)$$
Then, the Yang--Baxter equation (1.6) is completely equivalent to the
validity of the triple product equation;
$$\eqalign{[v,[u,&e_j,z]_{\theta^\prime} , [e^j, x, y]_\theta
 ]^*_{\theta^{\prime \prime}}\cr
&= [u, [v,e_j,x]^*_{\theta^\prime} , [e^j, z, y]^*_{\theta^{\prime \prime}}
]_\theta\cr}\eqno(6.2)$$
without any additional ansatz.
Note that Eq. (6.2) is invariant under
$$u \leftrightarrow v,\quad  x \leftrightarrow z,
\quad \theta \leftrightarrow \theta^{\prime \prime},\quad
 [x,y,z] \leftrightarrow
[x,y,z]^*\quad . \eqno(6.3)$$
  Especially, we need not even assume that
the inner product $<x \vert y>$ is either symmetric or antisymmetric as we
have done in the present note.  If $[x,y,z]^*_\theta =
[x,y,z]_\theta$, then Eq. (6.2) reduces to Eq. (1.8).  We will
 make an attempt to
solve the general equation (6.2) in the future with possible uses of more
general triple systems other than OTS and STS considered in this note.

Finally, it may be worthwhile to briefly sketch a history of uses of triple
products in theoretical physics.  It appears that Y. Nambu$^{19)}$ was the
first person to have suggested a possible generalization of Heisenberg
equation of motions in the quantum mechanics by introducing some triple
products.  Also, I. Bars$^{20)}$ has attempted to use triple systems to be
somehow related to sub--constituent blocks of quarks and leptons.  On the
other side, Truini and Biedenharn$^{21)}$ have utilized the so--called
Jordan--pair system (which is somewhat related to our Freudenthal's triple
system) in their model of constructing a grand--unified theory.  More
recently, G\"unaydin and co--workers$^{22)}$ have works utilizing ternary
algebras for constructions of superconformal algebras.
\medskip
\noindent {\bf \underbar{Acknowledgement}}

This work is supported in part by the U.S. Department of Energy Grant
No.

\noindent DE--FG02--91--ER40685.
\vfil\eject
\noindent {\bf \underbar{References}}
\item{1.} S. Okubo; University of Rochester Report UR--1293 (1992).
\item{2.} K. Yamaguchi and H. Asano; Proc. Jap. Acad. {\bf 51}, 247 (1972).
\item{3.} J. R. Faulkner and J. C. Ferrar, Indago Math {\bf 34},
 247 (1972).
\item{4.} I. L. Kantor; Sov. Math. Dokl. {\bf 14}, 254 (1973).
\item{5.} W. Hein; Trans. Amer. Math. Soc. {\bf 205}, 79 (1975), Math.
 Ann. {\bf 213}, 195 (1975).
\item{6.} B. N. Allison; Amer. J. Math. {\bf 98}, 285 (1976).
\item{7.} I. Bars and M. G\"unaydin; J. Math. Phys. {\bf 20}, 1977 (1979).
\item{8.} Y. Kakiichi; Proc. Jap. Acad. {\bf 57}, Ser A. 276 (1981).
\item{9.} K. Yamaguchi; Bull. Fac. Sch. Ed. Hiroshima University
{\bf 6} (2), 49 (1983).
\item{  } K. Yamaguchi and A. Ono; ibid, Part II, {\bf 7}, 43 (1984).
\item{10.} W. G. Lister; Amer. J. Math. {\bf 89}, 787 (1952).
\item{   } K. Yamaguchi; J. Sci. Hiroshima University {\bf A21}, 155
(1958).
\item{11.} N. Kamiya; Mem. Fac. Sci. Shimane University {\bf 22}, 51
 (1988).
\item{12.} H. Freudenthal; Indago Math {\bf 16}, 218 (1954), 363 (1954),
{\bf 21}, 447 (1959), {\bf 25}, 457 (1963).
\item{13.} H. J. de Vega and H. Nicolai; Phys. Lett. {\bf B244}, 295
 (1990).
\item{14.} A. B. Zamolodchikov and A$\ell$. B. Zamolodchikov,
 Nucl. Phys. {\bf B133}, 525 (1978).
\item{15.} S. Okubo; Alg. Group Geom. {\bf 3}, 60 (1986).
\item{16.} H. Weyl; \lq\lq Classical Group"  Princeton University Press,
Princeton, NJ (1939).
\item{   } D. E. Littlewood, \lq\lq The Theory of Group
Character", Clarendon, Oxford (1940).
\item{17.} H. Asano; \lq\lq Symplectic Triple Systems and Simple Lie
Algebras" in
S$\overline{\rm u}$rikagaku K$\overline{\rm o}$ky$\overline{\rm u}$roku
308, University of Kyoto, Inst. Math. Analysis (1977) in Japanese.
\item{18.} e.g. J. E. Humphreys; \lq\lq Introduction to Lie Algebras and
Representation Theory" Springer--Verlag, New York--Heidelberg--Berlin, 3rd
edition (1980).
\item{19.} Y. Nambu; Phys. Rev. {\bf D7}, 2405 (1973).
\item{20.} I. Bars; in Proceeding of the IX--th International Colloquium on
Group Theoretical Method in Physics, Cocoyoc, Mexico, 259 (1980).
\item{21.} P. Truini and L. C. Biedenharn; J. Math. Phys. {\bf 23}, 1327
(1982).
\item{22.} M. G\"unaydin; Phys. Lett. {\bf B255}, 46 (1991).
\item{   } M. G\"unaydin and S. Hyun; Mod. Phys. Lett. {\bf 6}, 1733
(1991), Nucl. Phys. {\bf B373}, 688 (1992).
\item{   } M. G\"unaydin; Penn. State University Report, PSU--TH--107
(May, 1992) unpublished.

\bye